\documentstyle[amscd,amssymb,verbatim,12pt]{amsart}
\theoremstyle{plain}

\theoremstyle{definition}

\numberwithin{equation}{section}

\def\dspace{\baselineskip=0.3 in}
\begin{document}
\dspace
\title[Quintessence Dark energy ...]{Quintessence Dark Energy inspired by Dual Roles of the Ricci Scalar }

\author[S.K.Srivastava]%
        {    }

\maketitle

\centerline{\bf S.K.Srivastava }

\centerline{ Department of Mathematics, North Eastern Hill University,}

\centerline{ NEHU Campus,Shillong - 793022 ( INDIA ) }

\centerline{e-mail:srivastava@@nehu.ac.in ; sushil@@iucaa.ernet.in }

\smallskip

\centerline{\bf Abstract}

\smallskip

In this letter, dark energy is obtained using dual roles of the Ricci scalar
(physical as well as geometrical role of the Ricci scalar curvature) in a
natural way without any input in the theory for it. In contrast to other
 models, where an idea for dark energy is introduced in the beginning of the
 theory, here it emerges from the gravitational sector
 spontaneously. Dark energy density, obtained  in this model,
decreases with the scale factor $a(t)$ as $\sim a(t)^{-3 (1 + {\rm w}_{\rm
    (de)})}$ (where ${\rm w}_{\rm (de)} > - 1$ is the equation of state parameter for dark
energy ) explaining fall
of its value from $1.19\times 10^{75}{\rm GeV}^4$ at Planck scale to its
present value suggested by astronomical observations. Interestingly, apart from dark
energy, dark radiation and dark matter  density are  also obtained from the gravity sector. Moreover, time for transition
from deceleration to acceleration of the universe is evaluated. It is found
that if dark energy obeys equation of state for generalized Chaplygin gas and
barotropic fluid simultaneously, after tansition time, the scale factor will
evolve as $cosh (t - t_*) (t$ and $t_*$ being time and transition time
respectively), but after a finite time expansion will become de-Siiter like. 

\vspace{1cm}

\noindent 1. Astronomical observations, during the last few years, provide compelling
evidences in favor of accelerating universe at present, which is caused by
dominance  of dark energy (DE) \cite{sp,ar,na,dn,rs}. Observation
of 16 Type Ia supernovae (SNe Ia) by $Hubble$ $Space$ $Telescope$ further modifies
these results and shows evidence for cosmic deceleration preceding 
acceleration in the late universe \cite{ag}. So, DE has a significant role in cosmic dynamics of
the current universe. The simplest candidate for DE is supposed to be the
cosmological constant $\Lambda$ which is very high in the early universe, but
there is no mechanism to bring it to present value without
fine-tuning. Alternatively, to explain its decay from a very high value, in
the early universe, to its present extremely small value, many models
\cite{jm,vs} were suggested in which $\Lambda$ is envisaged as a slowly
varying function of cosmic time. Apart from dynamical $\Lambda$ as DE, other
models are fluid-dynamics models, where barotropic fluid is its source with or
without dissipative pressure , Chaplygin gas(GC) and generalized Chaplygin gas
(GCG)models \cite{rj, ob,mc,sks}. In
field-theoretic models, the most natural ones are models, where DE is caused
by scalars. These are quintessence models \cite{brp}, k-essence models
\cite{ca}, tachyon models \cite{as} and phantom models \cite{sks, rr}. In
these models, non-gravitational lagrangian density for exotic matter giving DE
is added to Einstein-Hilbert term in the action. Recently, in a different
appraoch, non-gravitational term is replaced by gravitational term being
non-linear in Ricci scalar $R$, which stems modified gravity
\cite{sc,sm,add,sno,sns,de,smc,scv,snsd, ms,ka,om}.

In all these models, a non-gravitational or gravitational term for DE is added
in the theory {\it a priori}. In this sense, these models are phenomenological. 
 Cosmology demands a DE model, where DE density $\rho_{\rm (de)}$
emerges spontaneously from a basic theory. With this motivation,
here,$\rho_{\rm (de)}$ is derived from the gravitational sector without an
input for it. Here also, a quardatic term of R is added to  Einstein-Hilbert
term in the action, but not as lagrangian density for DE. It is unlike the
approach in \cite{sc,sm,add,sno,sns,de,smc,scv,snsd,ms,ka,om}. In the pesent model, DE emerges
spontaneously due to presence of term quardatic in R manifesting dual role of
the Ricci scalar as a physical field as well as geometry.

Natural units ($\hbar = c = 1$) are used here with GeV as the fundamental
unit, where $\hbar$ and $c$ have their usual meaning. In this unit, it is
found that $1 {\rm GeV}^{-1} = 6.58 \times 10^{-25} {\rm sec}$.

\bigskip

\noindent 2.The action
for higher-derivative gravity is taken as 
$$ S_g = \int {d^4x} \sqrt{- g} \Big[ \frac{R}{16 \pi G} - \alpha(x) R^2
\Big], \eqno(1)$$
where $R$ is the Ricci scalar curvature, $G = M_P^{-2} ( M_P = 10^{19}$ GeV is
the Planck mass). Moreover, dimensionless $\alpha $ is a scalar depending
space-time coordintaes.

The action (1) yields gravitational field equations
$$\frac{1}{16 \pi G} ( R_{\mu\nu} - \frac{1}{2} g_{\mu\nu} R ) - \alpha  ( 2
\triangledown_{\mu} \triangledown_{\nu}R - 2 g_{\mu\nu} {\Box} R - \frac{1}{2} g_{\mu\nu} R^2 + 2 R
R_{\mu\nu} )$$
$$ - 2 (\triangledown_{\mu} \triangledown_{\nu} \alpha -  g_{\mu\nu}
{\Box}\alpha ) R - 4 (\triangledown_{\mu} \alpha \triangledown_{\nu}R -
g_{\mu\nu} \triangledown^{\sigma} \alpha \triangledown_{\sigma} R ) = 0 \eqno(2)$$
using the condition $\delta S_g/{\delta g^{\mu\nu}} = 0.$ Here, $\triangledown_{\mu}$
denotes covariant derivative and the operator $\Box$ is given as
$${\Box} = \frac{1}{\sqrt{-g}} \frac{\partial}{\partial x^{\mu}}
\Big(\sqrt{-g} g^{\mu\nu} \frac{\partial}{\partial x^{\nu}} \Big) \eqno(3)$$
with $\mu, \nu = 0,1,2,3$ and $g_{\mu\nu}$ as metric tensor components.

Taking trace of eqs.(2), it is obtained that
$${\Box}R  + m^2 R + \frac{2}{\alpha}\triangledown^{\nu} \alpha
\triangledown_{\nu}R = 0 , \eqno(4a)$$
where
$$ m^2 = \frac{1}{96 \pi G \alpha (t) } + \frac{{\Box}\alpha}{\alpha} \eqno(4b)$$
with $ \alpha  \ne 0$ to avoid the ghost problem.  Here, overdot gives derivative with respect to time $t$.

Eq.(4a) shows that the Ricci scalar $R$ behaves as a physical field also with
(mass)$^2$ depending on $G$, in addition to
its usual role as a geometrical field \cite{ks, aas, skp}.

Experimental evidences support spatially homogeneous flat model of the
universe \cite{ad}. So, the line-element, giving geometry of the universe, is
taken as
$$ dS^2 = dt^2 - a^2(t)[dx^2 + dy^2 + dz^2] \eqno(5)$$
with $a(t)$ as the scale factor.

In the space-time, given by eq.(5), eq.(4a) is obtained as
$$ {\ddot R} + \Big(3 \frac{\dot a}{a} + 2\frac{\dot
  \alpha}{\alpha} \Big) {\dot R} - \Big\{\frac{1}{96 \pi G \alpha (t) } +
  \frac{\ddot \alpha}{\alpha} + 3 \frac{\dot a}{a}\frac{\dot
  \alpha}{\alpha}\Big\} R = 0, \eqno(6)$$

In most of the situations, for example, radiation model, matter-dominated
model, and accelerated models, we have $a(t)$ as a power-law solution yielding
$R$ as the power-law function of $a(t)$ . So, it is reasonable to take $R$
as
$$ R = \frac{A}{R^n}  \eqno(7)$$
with $n \ne 0$ being a real number. Moreover, $\alpha$ is taken as

$$ \alpha = D a^r, \eqno(8)$$
where $r$ is a non-zero real number. Advantage for taking $\alpha$ in this
form is mentioned below. Here $A$ and $D$ are constants.

$R$ and $\alpha$, given by eqs.(7) and (8) respectively, satisfy eq.(6), if
$$ \frac{\ddot a}{a} + (2 + r - n) \Big(\frac{\dot a}{a} \Big)^2 = - \frac{1}{96
  \pi G D (n - r) a^r }, \eqno(9)$$
 which integrates to
$$ \Big(\frac{\dot a}{a} \Big)^2 = \frac{C}{a^{2[3 - (n - r)]}} - \frac{1}{48
  \pi G D (n - r) [(6 - r) - 2(n - r)] a^r } \eqno(10)$$
with $C$ being an integration constant.
Eq.(10) gives dynamics of the universe, which is the Friedmann equation
$$\Big(\frac{\dot a}{a} \Big)^2 = \frac{8\pi G}{3} ( \rho_{\rm   de} + \rho_{r(m)})  \eqno(11)$$
with
$$\rho_{\rm   de} = \frac{3 C}{8\pi G a^{2[3 - (n - r)]}} \eqno(12a)$$
and
$$\rho_{r(m)} = \frac{3}{8\pi G} \Big[- \frac{1}{48
  \pi G D (n - r) [(6 - r) - 2(n - r)] a^r } \Big] .\eqno(12b)$$

It is interesting to note that $\alpha(t) \propto a^r$, gives energy density
in the known form $\propto a^{-r}$. 

$\rho_{r(m)}$, given by eq.(12b), takes the known form of energy density in
two cases (i) $ r = 4$ and (ii) $ r = 3$. When $ r = 4$, it
takes the form of radiation density $\rho_r \sim a^{-4}.$ It is in the form of
matter (pressureless fluid) density $\rho_m \sim a^{-3},$ when $ r = 3$.

$\rho_{\rm de}$, given by (12a), emerges from the gravitational sector without
  using a source for it. So,it is recognized as DE density.

Conservation equation for DE 
$$ {\dot \rho_{\rm (de)}} + 3 \frac{\dot a}{a} \rho_{\rm (de)}(1 + {\rm w}_{\rm
    (de)}) = 0,   \eqno(13)$$
where equation of state (EOS) parameter ${\rm w}_{\rm  (de)} = p_{\rm  (de)}/\rho_{\rm (de)}$ with $p_{\rm
  (de)}$ and $\rho_{\rm (de)}$ being isotropic pressure and density for DE
    respectively. Eq.(13) yields
$$ 2[3 - (n - r)] = 3 (1 + {\rm w}_{\rm  (de)}) , \eqno(14)$$ 

Thus, eqs.(12a) and (14) imply
$$\rho_{\rm   de} = \frac{3 C}{8\pi G a^{3 (1 + {\rm w}_{\rm  (de)})} }
  .\eqno(15)$$

 WMAP data give the present value of DE to be $\rho^0_{\rm (de)} = 0.73
 \rho_{\rm cr.},$ where $\rho_{\rm cr.} = 3 H_0^2/ 8 \pi G$ with $H_0 =  100 h
 {\rm km/sec Mpc} = 2.33 \times 10^{-42} {\rm GeV}$ and $h = 0.68$ (having
 maximum likelihood) \cite{abl,am03}. Using these values, eq.(17) is obtained as
$$ \rho_{\rm   de} = 0.73 \rho_{\rm cr.} \Big(\frac{a_0}{a}\Big)^{3 (1 + {\rm
    w}_{\rm  (de)})}. \eqno(16)$$

At Planck scale, $\rho_{\rm (de)} = \frac{3}{8 \pi G} M_P^2$, so eq.(24)
    yields
$$\Big(\frac{a_0}{a_P}\Big)^{3 (1 + {\rm
    w}_{\rm  (de)})} = \frac{M_P^2 H_0^{-2}}{0.73} = 5.46 \times 10^{121} , \eqno(17)$$
where $a_P$ is the scale factor at Planck time $t_P = M_P^{-1}$.
\smallskip

2(a). When $r = 4$, eq.(12b) is obtained as
$$ \rho_r = \frac{3 }{8\pi G } \Big[\frac{1}{72 \pi G D (1 - {\rm w}_{\rm de})
  |-1 + 3{\rm w}_{\rm de}| a^4} \Big] , \eqno(18)$$
which looks like energy density for radiation emerging from gravity sector. This type of term also arises
in brane-gravity inspired Friedmann equation, which is called as `dark energy
density' \cite{rm}. So, like brane-gravity, here also
$ \rho_r$ (given by eq.(18)) is termed as `dark radiation
density'.

Connecting eqs.(11), (12a) and (16), it is obtained that
$$\Big(\frac{\dot a}{a} \Big)^2 =  \frac{E}{ a^4} + 0.73 H^2_0
    \Big(\frac{a_0}{a}\Big)^{3 (1 + {\rm  w}_{\rm  (de)})} \eqno(19)$$
with $E = 1/\{72 \pi G D (1 - {\rm w}_{\rm de})
  |-1 + 3{\rm w}_{\rm de}| \}$.

For $\rho_r > \rho_{\rm (de)}$ , eq.(19) reduces to
$$ \Big(\frac{\dot a}{a} \Big)^2 = \frac{E}{a^4} , \eqno(20)$$
which integrates to
$$ a^2 = a^2(0) + 2 t \sqrt{E} \eqno(21)$$
with $a(0) = a(t=0).$ It shows deceleration as ${\ddot a} < 0.$

Investigations start here from the Planck scale (being the fundamental
scale).Using $a_P = a(t=t_P)$ with Planck time $t_P = M_P^{-1}$ in
eq.(18), $C$ is evaluated as
$$ C = \frac{1}{4} (a^2_P - a^2(0))^2 M_P^2 \simeq \frac{1}{4} a^4_P
M_P^2. \eqno(22)$$
as $a(0) < a_P.$ Thus,
$$ a^2 \simeq a^2_P M_P t. \eqno(23)$$

Connecting eqs.(16), (17), (18) and (22), it is obtained that $\rho_r >
\rho_{\rm de}$ if
$$ \Big(\frac{a_0}{a} \Big)^{1 - 3 {\rm w}_{\rm de}} > 5.15 \times
10^{-112}\times (5.46\times 10^{121})^{4/3(1 + {\rm w}_{\rm de})} . \eqno(24)$$

For ${\rm w}_{\rm de} = - 0.89,$ this inequality yields
$$ 1 + z = \frac{a_0}{a} > 10^{365} , \eqno(25)$$
where $z$ is the red-shift.

It is found that universe decelerates when  $\rho_r >
\rho_{\rm de}$. Moreover, universe accelerates when DE dominates. It means
that transition from dceleration to acceleration is possible when transition
from  $\rho_r > \rho_{\rm de}$ to  $\rho_r < \rho_{\rm de}$ takes
place. According to (25), it is possible at red-shift $10^{365}$, which is
extremely large. 16 Type Supernova suggest red-shift for this transition as $z
= 0.46 \pm 0.13$ \cite{ag}. Thus this case (for $(n - r) = 1$) contradicts
experimental results. So, it is ignored here onwards.  

\smallskip

2(b). When $ r = 3$, eq.(12b) looks like
$$ \rho_m = \frac{3 B}{8\pi G a^3} \eqno(26)$$
with $B = \frac{1}{216 \pi G (1 - {\rm w}_{\rm de})|{\rm w}_{\rm de}|}.$  
The energy density (26) has the form of density for pressueless matter
emerging from gravity. So, it is termed as dark matter density (DMD). The
prtesent value of DMD is $\rho^0_m = 0.23 \rho_{\rm cr}$ (critical density
$\rho_{\rm cr}$ is defined above). So, (26) is obtained as
$$ \rho_m = 0.23 \rho_{\rm cr} \Big(\frac{a_0}{a} \Big)^3 . \eqno(27)$$

Now, $\rho_m > \rho_{\rm de}$ when red-shift $z$ is given as
$$ z > \Big(\frac{73}{23} \Big)^{1/3|{\rm w}_{\rm de}|} - 1. \eqno(28)$$

 $\rho_m < \rho_{\rm de}$  for
$$ z < \Big(\frac{73}{23} \Big)^{1/3|{\rm w}_{\rm de}|} - 1. \eqno(29)$$

Inequalities (28) and (29) show that transition from $\rho_m > \rho_{\rm de}$
to $\rho_m < \rho_{\rm de}$ at
$$z_* = \Big(\frac{73}{23} \Big)^{1/3|{\rm w}_{\rm de}|} - 1 =
\Big(\frac{a_0}{a_*} \Big) - 1. \eqno(30)$$ 

Observations yield $-1 < {\rm w}_{\rm de} < - 0.82$ for quintessence DE. Using these values of ${\rm w}_{\rm de}$ , (30) yields
$$ 0.47 < z_* \lesssim 0.599   ,  \eqno(31)$$
which is supported by the range $ 0.33 \lesssim  z_* \lesssim 0.59$ given by
16 Type Supernova observations.

Using (16) and (27), Friedmann equation (11) is obtained as

$$\Big(\frac{\dot a}{a} \Big)^2 = H_0^2 \Big[0.23 \Big(\frac{a_0}{a} \Big)^3 +
0.73 \Big(\frac{a_0}{a} \Big)^{3(1 + {\rm w}_{\rm de})} \Big] \eqno(32)$$
using definition of $\rho^0_{\rm cr}$ given above.

When $z > z_*, \rho_m > \rho_{\rm de}$, so (32) reduces to
$$\Big(\frac{\dot a}{a} \Big)^2 = 0.23 H_0^2 \Big(\frac{a_0}{a}
\Big)^3,  \eqno(33)$$ 
which integrates to
$$ a(t) = a_d \Big[ 1 + 0.72 H_0 \Big(\frac{a_0}{a_d} \Big)^{3/2} (t - t_d)
\Big]^{2/3}  \eqno(34)$$
with $a_d = a(t_d)$ being a constant,$ a(t) $, given by (34), shows deceleration.

Further, for  $z < z_*, \rho_m < \rho_{\rm de}$, so (32) reduces to
$$\Big(\frac{\dot a}{a} \Big)^2 = 0.73 H_0^2 \Big(\frac{a_0}{a} \Big)^{3(1 +
  {\rm w}_{\rm de})} ,  \eqno(35)$$ 
which is integrated to
$$ a(t) = a_* \Big[ 1 +  \frac{3(1 + {\rm w}_{\rm de})}{2} H_0 \sqrt{0.73}
  \Big(\frac{a_0}{a_*} \Big)^{3(1 +   {\rm w}_{\rm de})/2} (t - t_*) \Big]^{2/3(1 + {\rm w}_{\rm de})}   \eqno(36)$$     
with $a_* = a(t_*)$. (36) shows acceleration. Here $t_*$ is the time for
  transition from deceleration to acceleration. Using $a_0 = a(t_0)$ in (36),
  we obtain that
$$ t_0 - t_* = \Big[ 1 +  \frac{3(1 + {\rm w}_{\rm de})}{2} H_0 \sqrt{0.73}
\Big]^{-1} \Big[1 - \Big(\frac{a_*}{a_0} \Big)^{3(1 +   {\rm w}_{\rm de})/2}
\Big] . \eqno(37)$$

Connecting (30) and (37), $t_*$ is evaluated as
$$ t_* = 0.627 t_0 = 8.59 {\rm Gyr}  \eqno(38)$$
for ${\rm w}_{\rm de} = - 0.89$ and present age of the universe $t_0 = 13.7
{\rm Gyr}$ \cite{abl}.

\bigskip

\noindent 3. So far, we have considered DE (obtained here) as a barotropic
fluid obeying EOS
$$ p_{\rm de} = {\rm w}_{\rm de} \rho_{\rm de} . \eqno(39a)$$

Due to negative pressure and being only fluid with supersymmetric
generalization , GC and GCG have been strong candidates for DE. But
experimental results support GCG \cite{ob,mc}. So, GCG model is preferred.

In what follows, consequences of DE fluid (obtained here) are
explored if it behaves as GCG
\cite{rj,ob,mc,sks} and barotropic fluid simultaneously. This type of
situation is explored in \cite{sks} for non-gravitational phantom DE. The equation of state (EOS) for GCG is given as
$$ p_{\rm de} = - \frac{M^{1 + \beta}}{\rho_{\rm de}^{\beta}}, \eqno(39b)$$  
where $0 < \beta < 1$ for GCG , $\beta = 1$ for CG and $M$ is a constant.

(39a) and (39b) yield
$$ {\rm w}_{\rm de} = - \Big(\frac{M}{\rho_{\rm de}} \Big)^{1 +
  \beta}. \eqno(40)$$ 
This equation shows dependence of $ {\rm w}_{\rm de}$ on $\rho_{\rm de}$
  varying with $a(t)$. Now, using eq.(16) for $\rho_{\rm de}$ with variable $
  {\rm w}_{\rm de}$, conservation equation (13) looks like
$$3 {\dot {\rm w}_{\rm de}} + \frac{3}{lna} \Big( \frac{\dot a}{a } \Big) (1 + {\rm
  w}_{\rm de}) = \frac{3}{lna} \Big( \frac{\dot a}{a } \Big)
 \Big[ 1 - \Big(\frac{M}{B} \Big)^{1 + \beta} e^{3(1 +
  \beta)(1 + {\rm w}_{\rm de})ln a} \Big] \eqno(41a)$$ 
using eq.(40) for ${\rm w}_{\rm de}.$ Here,
$$ B = 0.73 \rho_{\rm cr} a_0^{3 (1 + {\rm w}_{\rm de})} . \eqno(41b)$$
Eq.(41a) integrates to
$$ a^{ - 3(1 +   \beta)(1 + {\rm w}_{\rm de})} = \frac{\tilde C}{a^{3(1 +
    \beta)}} +  \Big(\frac{M}{B} \Big)^{1 + \beta}, \eqno(42)$$
where $\tilde C$ is an integration constant.

So, eqs.(16), (41b) and (42) yield
$$ \rho_{\rm de} = B \Big[\frac{\tilde C}{a^{3(1 +
    \beta)}} +  \Big(\frac{M}{B} \Big)^{1 + \beta} \Big]^{1/(1 + \beta)} . \eqno(43)$$
Connecting eqs.(40) and (43), it is obtained that
$$ {\rm w}_{\rm de} = - \Big(\frac{M}{B} \Big)^{1 + \beta}\Big[\frac{\tilde C}{a^{3(1 +
    \beta)}} +  \Big(\frac{M}{B} \Big)^{1 + \beta} \Big]^{-1} . \eqno(44)$$ 

At $a = a_0, {\rm w}_{\rm de} = {\rm w}^0_{\rm de},$ so eq.(44) yields
$$ - \Big(\frac{M}{B} \Big)^{1 + \beta} = \Big[\Big(\frac{{\rm w}^0_{\rm de}}{1 +
  {\rm w}^0_{\rm de}} \Big) \frac{\tilde C}{a_0^{3(1 + \beta)}} - 1
  \Big]^{-1}. \eqno(45a)$$
Connecting eqs.(44) and (45a), we obtain
$${\rm w}_{\rm de} = \Big[\Big(\frac{1 +   {\rm w}^0_{\rm de}}{{\rm w}^0_{\rm
      de}}\Big) \Big(\frac{a_0}{a} \Big)^{3(1 + \beta)} - 1
      \Big]^{-1}. \eqno(45b)$$ 
Eqs.(43) and (45a) yield
$$ \rho_{\rm de} = \frac{B {\tilde C}^{1/(1 + \beta)}}{a_0^3} \Big[
\Big(\frac{a_0}{a} \Big)^{3(1 + \beta)} - \Big(\frac{{\rm w}^0_{\rm de}}{1 +
  {\rm w}^0_{\rm de}} \Big) \Big]^{1/(1 + \beta)}. \eqno(46a)$$

At $a = a_0, \rho_{\rm de} = \rho^0_{\rm de} = 0.73 \rho_{\rm cr}, $ so
$$ \rho_{\rm de} = 0.73 \rho_{\rm cr} \Big[ (1 +
  {\rm w}^0_{\rm de})\Big(\frac{a_0}{a} \Big)^{3(1 + \beta)} - {\rm w}^0_{\rm
  de}  \Big]^{1/(1 + \beta)} \eqno(46b)$$
setting $\tilde C = a^{-3 {\rm w}^0_{\rm de}(1 + \beta)}.$

The constant M, in (40), is evaluated as
$$ M^{1 + \beta} = - {\rm w}^0_{\rm de} (\rho^0_{\rm de})^{1 + \beta}  \eqno(47)$$
using ${\rm w}^0_{\rm de} = {\rm w}_{\rm de}(t_0).$ So, connecting (40), (46b)
and (47), it is obtained that
$${\rm w}_{\rm de} = {\rm w}^0_{\rm de} \Big[ (1 + {\rm w}^0_{\rm de})
\Big(\frac{a_0}{a} \Big)^{3(1 + \beta)} - {\rm w}^0_{\rm de}\Big]^{-1} . \eqno(48)$$

In the changed circumstances,  $\rho_m < \rho_{\rm de}$ when
$$ \frac{23}{73} \Big(\frac{a_0}{a} \Big)^3 < \Big[ (1 + {\rm w}^0_{\rm de})
\Big(\frac{a_0}{a} \Big)^{3(1 + \beta)} - {\rm w}^0_{\rm de}\Big]^{1/(1 + \beta)} . \eqno(49)$$
This inequality is obtained using (27) and (46b). It shows that, for GCG ,
transition from $\rho_m > \rho_{\rm de}$ to $\rho_m < \rho_{\rm de}$ may take
place when
$$\Big(\frac{a_0}{a_*} \Big)^{3(1 + \beta)}_{\rm GCG} = -  {\rm w}^0_{\rm de}
\Big[ \Big(\frac{23}{73} \Big)^{1 + \beta} - (1 + {\rm w}^0_{\rm de})
\Big]^{-1} .  \eqno(50)$$

Taking ${\rm w}^0_{\rm de} = - 0.89$ in (50),
$$ z_{*{\rm (GCG)}} = \Big(\frac{a_0}{a_*} \Big) - 1 = 0.64  \eqno(51)$$
for $\beta = 0.04$. This value of red-shift for transition is very near to
upper limit $0.59$ of $z$, given in \cite{ag}. So, in GCG case, $\rho_m <
\rho_{\rm de}$ when $z < z_* = 0.64$. Now, using (46b) and definition of
$\rho_{\rm cr}$, Friedmann Equation looks like
\begin{eqnarray*}
\Big(\frac{\dot a}{a} \Big)^2 &=&  0.73 H_0^2 \Big[ (1 +
  {\rm w}^0_{\rm de})\Big(\frac{a_0}{a} \Big)^{3(1 + \beta)} - {\rm w}^0_{\rm
  de}  \Big]^{1/(1 + \beta)} \\& \simeq & 0.73 H_0^2 |{\rm w}^0_{\rm
  de}|^{1/(1 + \beta)} \Big[ \Big\{a^{3(1 + \beta)} + \frac{(1 +
  {\rm w}^0_{\rm de})}{2 |{\rm w}^0_{\rm   de}|(1 + \beta)}a_0^{3(1 + \beta)}
  \Big\}^2 \\ &&- \frac{(1 + 2 \beta)}{4 (1 + \beta)^2} \Big(\frac{(1 +
  {\rm w}^0_{\rm de})}{ |{\rm w}^0_{\rm   de}|} \Big)^2 a_0^{6(1 + \beta)}
  \Big] ,
\end{eqnarray*}
$$ \eqno(52)$$
which is integrated to
$$ a(t) = a_{*{\rm (GCG)}} \Big[\frac{\sqrt{1 + 2 \beta} cosh \theta -
  1}{\sqrt{1 + 2 \beta}  - 1} \Big]^{1/3(1 +  \beta)},  \eqno(53a)$$
where
$$ \theta = 3 \sqrt{0.73} H_0 (1 +  \beta) |{\rm w}^0_{\rm   de}|^{1/2(1 +
  \beta)} (t - t_{*{\rm (GCG)}}) .  \eqno(53b)$$

Using $\beta = 0.04$ and $z_* = 0.64, t_{*{\rm (GCG)}}$ is evaluated as
$$ t_{*{\rm (GCG)}} \simeq 0.8 t_0 = 10.9 {\rm Gyr}  .  \eqno(54)$$   

Connecting (53a) and (53b), it is obtained that
$$\frac{\dot a}{a} = \sqrt{0.73} H_0 |{\rm w}^0_{\rm   de}|^{1/2(1 +
  \beta)}\Big[\frac{\sqrt{1 + 2 \beta} tanh \theta }{\sqrt{1 + 2 \beta}  -
  sech \theta} \Big] . \eqno(55a)$$
It shows that
$$\frac{\dot a}{a} \simeq \sqrt{0.73} H_0 |{\rm w}^0_{\rm   de}|^{1/2(1 +
  \beta)}  \eqno(55b)$$
for $\theta = 12.2$. Using $\theta$, from (53b), The  corresponding time
  $t_{\rm ds}$, for $\theta = 12.2$, is obtained as
$$ t_{\rm ds} = t_{*{\rm (GCG)}} + \frac{4.76 H_0^{-1}}{(1 +
  \beta)}|{\rm w}^0_{\rm   de}|^{- 1/2(1 + \beta)} . \eqno(56)$$
Using $\beta = 0.04$ and ${\rm w}^0_{\rm   de} = - 0.89$ in (50), $t_{\rm ds}$
  is evaluated as
$$ t_{\rm ds} = t_{*{\rm (GCG)}} + 4.88 t_0 \simeq 5.68 t_0 = 77.78 {\rm Gyr},
\eqno(57)$$  
where $t_{*{\rm (GCG)}}$ is given by (54).

For $t \ge t_{\rm ds}, tanh \theta = 1$ and $cosh \theta \simeq
\frac{e^{\theta}}{2}$ , so $a(t)$ (given by (53a)) looks like
$$ a(t) = a_{*{\rm (GCG)}}\Big[\frac{\sqrt{1 + 2 \beta} cosh \theta - 2
  }{\sqrt{1 + 2 \beta}  - 2} \Big]^{1/3(1 + \beta)} \simeq F e^{\sqrt{0.73}
  H_0 |{\rm w}^0_{\rm   de}|^{1/2(1 + \beta)}(t - t_{*{\rm (GCG)}})}  \eqno(58)$$
where $F = a_{*{\rm (GCG)}}\Big[\frac{\sqrt{1 + 2 \beta} }{\sqrt{1 + 2 \beta}
  - 2} \Big]^{1/3(1 + \beta)}$ with $\theta$, given by (53a). Moreover, for $t
  \ge t_{\rm ds}$,
$$ \rho_{\rm de(GCG)} \simeq 0.73 \rho_{\rm cr}|{\rm w}^0_{\rm   de}|^{1/(1 +
  \beta)} . \eqno(59)$$
(58) shows that ,in GCG case,when $t \gtrsim t_{\rm ds}$, expansion becomes
  de-Sitter like and and DE density acquires a constant value slightly less
  than present DE density. In non-GCG case, $\rho_{\rm de} \to 0$ as $t \to
  \infty$. From (48), it is found that $ - 0.63 > {\rm w}_{\rm   de} \gtrsim -
  1$ taking $\beta$ and ${\rm w}^0_{\rm   de}$ given above. But for $t \ge
  t_{\rm ds}, {\rm w}_{\rm   de} \simeq -1.$

Moreover, (53a) yields
$$ \Big(\frac{\ddot a}{a}\Big)_{\rm GCG} = 0.73 H_0^2 |{\rm w}^0_{\rm
  de}|^{1/2(1 + \beta)} \Big[1 - \frac{sech \theta}{\sqrt{1 + 2 \beta}} \Big]^{-1} . \eqno(60)$$ 
This equation shows that as $t$ increases from $t = t_{*{\rm (GCG)}},
  \Big(\frac{\ddot a}{a}\Big)_{\rm GCG}$ decreases and becomes $\simeq 0.73
  H_0^2 |{\rm w}^0_{\rm  de}|^{1/2(1 + \beta)}$ for $t \ge t_{\rm ds}$. It
  means that, in GCG case,  expansion is faster around transition time, but
  slows down as $t \to t_{\rm ds}$.

In non-GCG case, $a(t)$ is given by (36), which yields
$$ \Big(\frac{\ddot a}{a}\Big)_{\rm non-GCG} = 0.73 H_0^2 \Big[1 - \frac{3 (1
  + {\rm w}_{\rm  de})}{2} \Big] \Big(\frac{ a_0}{a}\Big)^{3 (1
  + {\rm w}_{\rm  de})}. \eqno(61)$$  

Thus, it is found that acceleration increases with growing scale factor $a(t)$
in both GCG and non-GCG cases. But, in GCG case, $\ddot a \propto a$ and
$\ddot a \propto a^{1 - 3 (1 + {\rm w}_{\rm  de})}$ for non-GCG case. It means
that, in GCG case, acceleration grows faster with increasing $a(t)$ compared
to the non-GCG case if ${\rm w}_{\rm  de} > -1$. Its reason is explained in the
following way.

Accelerated expansion is caused by DE, characterized by negative pressure
violating the {\sl strong energy condition}, which gives cosmic push due to
its reverse gravity effect. In non-GCG case, $-p$ decreases with increasing
scale factor (as $ -p \propto \rho$), but it increases with $a(t)$ for GCG case and attains a constant
value in finite time. So, GCG dark energy gives more cosmic push causing more
acceleration compared to non-GCG dark energy,

\bigskip

4.In this letter, it is found  dark energy emerges spontaneously from gravity sector using  dual nature of the Ricci scalar as a
geometrical field as well as physical field (mentioned above). Quintessence dark energy
density, obtained here, falls from very high value $1.19\times 10^{75} {\rm GeV}^4$ at the Planck
scale to its current value $0.73 \rho_{\rm cr.} \simeq 2.18 \times
10^{-47}{\rm GeV}^4 $. At the transition time $t_*$, which is  $8.59 {\rm   Gyr}$ for ${\rm   w}_{\rm de} = - 0.89$, dark energy begins to dominate
giving a sudden jerk to the universe. As a result, a transition from
deceleration to acceleration takes place showing the reverse gravity effect of
dark energy. It is also found that dark radiation is washed away in the late
  universe. Transition from dominance of DM to dominance of DE is obtained at
  red-shift $0.47 < z_* < 0.599$ (for $- 0.82 > {\rm   w}_{\rm de} > -1$) causing transition from deceleration to
  acceleration. Interestingly, $z_*$, obtained here, is consistent with the
  value of $z_*$ given by 16 Type SNe Ia observations. Universe accelerates after $t > t_*$ with power-law
expansion with vanishing energy density, when $t \to \infty$. But, if
DE fluid behaves as GCG and barotropic fluid simultaneously, accelerated
expansion is faster due to more cosmic push compared to non-GCG case. It is
found that, in GCG case, expansion becomes de-Sitter like after a finite time
with finite energy density.

\end{document}